# Optical signatures of silicon-vacancy spins in diamond

Tina Müller[1][*+], Christian Hepp[2+], Benjamin Pingault[1], Elke Neu[2,3], Stefan Gsell[4], Matthias Schreck[4], Hadwig Sternschulte[5,6], Doris Steinmüller-Nethl[5], Christoph Becher[2] and Mete Atatüre[1*]

[1]Cavendish Laboratory, University of Cambridge, JJ Thomson Avenue, Cambridge CB3 0HE, UK

[2]Fachrichtung 7.2 (Experimentalphysik), Universität des Saarlandes, Campus E2.6, 66123 Saarbrücken, Germany

[3]Departement Physik, Universität Basel, Klingelbergstrasse 82, 4056 Basel, Switzerland

[4]Experimentalphysik IV, Institut für Physik, Universität Augsburg, Universitätsstrasse 1 Nord, 86159 Augsburg, Germany

[5]KOMET RHOBEST GmbH, 6020 Innsbruck, Austria

[6]Fakultät für Physik, Technische Universität München, James-Franck-Strasse 1, 85748 Garching, Germany

[+]These authors contributed equally to this work.

[*]Corresponding authors: tm373@cam.ac.uk, ma424@cam.ac.uk

**Colour centres in diamond have emerged as versatile tools for solid-state quantum technologies ranging from quantum information [1-3] to metrology [4-7], where the nitrogen-vacancy centre is the most studied to-date. Recently, this toolbox has expanded to include different materials [8, 9] for their nanofabrication opportunities, and novel colour centres [10-16] to realize more efficient spin-photon quantum interfaces. Of these, the silicon-vacancy centre stands out with ultrabright single photon emission predominantly into the desirable zero-phonon line [14]. The challenge for utilizing this centre is to realise the hitherto elusive optical access to its electronic spin. Here, we report spin-tagged resonance fluorescence from the negatively charged silicon-**

vacancy centre. In low-strain bulk diamond spin-selective excitation under finite magnetic field reveals a spin-state purity approaching unity in the excited state. We also investigate the effect of strain on the centres in nanodiamonds and discuss how spin selectivity in the excited state remains accessible in this regime.

The silicon-vacancy centre in diamond consists of a silicon atom and a vacancy in a split vacancy configuration [17-19], replacing two neighbouring carbon atoms in a diamond matrix along the <111> axes [see Fig. 1 (a) inset]. Previous electron-spin resonance measurements on ensembles of neutral silicon-vacancy centres have identified an $S$=1 ground state [20-22]. The negatively charged silicon-vacancy (SiV⁻) centre should therefore be associated with an $S$=1/2 ground state – consistent with the number of electrons in the contributing chemical bonds [23].

At cryogenic temperatures the photoluminescence spectrum of the zero-phonon line reveals a characteristic fine structure composed of four transitions around 737 nm, as displayed in Fig. 1 (a) (transitions labelled from A to D). This spectral signature originates from the doubly split ground and excited states shown in Fig. 1 (b) [24, 25]. Magnetic field further lifts any degeneracies of these four transitions, as can be seen in Fig. 1 (c) for a single SiV⁻ centre in a nanocrystal (blue spectrum), and for an ensemble of SiV⁻ centres in bulk diamond (red spectrum). Common to both spectra is the measured quadruplet splitting in the optical transitions, which is consistent with an energy level scheme based on spin-1/2 ground and excited states [Fig. 1 (d)] [26]. Theoretical analysis based on density functional theory (DFT) [17], as confirmed by a recent *ab initio* study [27], has determined a $D_{3d}$ symmetry for the SiV⁻ centre and assigned the optical ground and excited states to $E_g$ and $E_u$ states, respectively. The system can then be described by a Hamiltonian comprising orbital and spin Zeeman terms, the Jahn-Teller effect, which partially lifts the orbital degeneracies and a spin-orbit coupling term ($\boldsymbol{L}\cdot\boldsymbol{S}$). Group theoretic analysis predicts that, in the case of the SiV⁻ centre, only the $L_zS_z$ component ($z$ along the $C_3$ axis of the centre) of the spin-orbit operator acts on the $E_g$ and $E_u$ states [28], providing an inherent quantization axis along the <111> directions but leaving the spin as a good quantum number analogous to the NV⁻ centres [29].

Resonance fluorescence through state-selective excitation is a powerful tool to reveal signatures of spin in quantum optics. We first study an ensemble of SiV⁻ centres in a low-strain bulk diamond at 0 T. Figure 2 (a) displays the SiV⁻ spectrum (red curve) observed when driving transition B resonantly. Here, the laser is suppressed by polarisation rejection (see Methods) and contributes a small fraction to the full spectrum [23]. Even though resonant excitation selectively populates one excited state branch, all four transitions are visible indicating a relaxation process between the excited state branches prior to photon emission. Tuning the laser frequency across transition A, while monitoring the fluorescence of transition C in the spirit of photoluminescence excitation, reveals the absorption profile of transition A [solid red circles in Fig. 2 (b)]. The extracted full width at half maximum (FWHM) of ~10 GHz is consistent with the inhomogeneous broadening of the ensemble under non-resonant excitation [30] due to residual strain in the sample. For a single centre in a nanodiamond, the fluorescence spectrum, obtained by driving transition A resonantly, is shown in Fig. 2(d) (blue curve). Due to the strain in the crystal, the transitions of this centre are shifted beyond the inhomogeneous broadening of the ensemble in bulk diamond, where the exact spectrum varies from centre to centre. The absorption linewidth of 1.4 GHz for this transition [Fig. 2 (e)] is only order of magnitude above the radiatively broadened limit (~100 MHz) [10, 24] which should be reachable straightforwardly using impurity-free diamond substrates, as was shown for NV centres [31]. This linewidth also marks the minimum Zeeman splitting needed to resolve spin sublevels spectrally under resonant excitation.

Applying a magnetic field of 4 T to the ensemble of SiV⁻ centres in bulk diamond allows us to address excited states with a given spin orientation selectively. First, we drive the transition labelled A2 [23] to populate a Zeeman sublevel of the upper branch of the excited state, expected theoretically to be a spin-up projection, as shown in Fig. 3 (a). The resulting spectrum, shown in Fig. 3 (b) (red shaded curve), is strikingly different from the spectrum obtained under non-resonant excitation [Fig. 3 (c)]: Only half of the available optical transitions are visible and they originate from two excited states only. This is in stark contrast to thermal distribution at 4 K, which would lead to a finite population probability for all excited states. Here, the relaxation process in the excited state takes place only between the two sublevels with the same Zeeman response, i.e. same spin projection. To populate an excited state sublevel with the opposite Zeeman response, expected theoretically to be a spin-down projection, we resonantly drive transition B3 [blue double arrow in Fig. 3 (d)]. The resulting spectrum, shown in Fig. 3(e) (blue shaded curve), is strongly anti-correlated with that in Fig. 3(b), and the sum of the two spectra produces the full spectrum observed under non-resonant excitation [Fig. 3(c)]. The spectra in Figs. 3 (b) and 3(e) further reveal that from the populated excited states, optical transitions to all ground states occur, irrespective of their spin projection. The anti-correlation in the spectra arises from the high degree of spin purity in the excited state. In contrast, what appears as a lack of selection rules in the optical transitions is a consequence of the orientation of the applied magnetic field affecting mostly the electronic ground state. Indeed, when the magnetic field is not parallel to the SiV⁻ axis (as is the case in our experiment) the original quantization axis of the centre is tilted due to the magnetic field induced non-diagonal terms in the spin-orbit basis, which in turn induces an apparent spin mixing. With an angle of 54.7° between the <111> SiV⁻ axes and the 4-T magnetic field along [001] the spin-orbit coupling strength in the excited state is still dominant and the spin-down projection sublevel in the higher lying excited state sustains a spin purity of 97%, which agrees well with the 4.1% ± 1.2 % overlap between the two anti-correlated spectra of Fig. 3 [23]. On the other hand, the spin-orbit coupling strength in the ground state is comparable to the off-axis contributions of the Zeeman Hamiltonian (at 4T) resulting in a degree of spin purity ranging from 50% to 80% for the ground-state branches [28]. This effective spin mixing gives rise to the observed relaxation of the optical selection rules. Aligning the magnetic field along the $C_3$ axis of the SiV$^-$ centre is expected to restore more than 90% spin purity in the ground state and near unity spin purity in the excited state. These properties mark the SiV$^-$ centre desirable for all applications that require optical access to well-defined spin states.

We now investigate the influence of strain through resonance fluorescence from a single centre located in a nanodiamond. A magnetic field of 2 T allows us to optically resolve the individual transitions of the centre shown in Fig. 1 (c). The excitation laser is brought into resonance with transition A1 [as shown in Fig. 4 (a)] and transition A2. This leads to fluorescence spectra with selective population of the spin-up (red curve) and spin-down (blue curve) sublevels in the higher branch of the excited state, respectively, as seen in Fig. 4(b). The signature of spin selectivity demonstrated in bulk is only partially observed in the resulting fluorescence spectra shown in Fig. 4 (b). While the transitions B1 - B4, originating directly from the higher-lying branch of the excited state, still exhibit a high degree of spin selectivity (hence spin purity), the transitions C1 - D4, which originate from the lower branch after a relaxation step, do not display such spin selectivity. This breakdown is most likely to be induced by the strong strain field not oriented with the symmetry axis of the centre, however further investigation is required to understand the full origin of this mechanism.

Bypassing the inter-branch relaxation mechanism by directly populating a spin sublevel of the lower branch, allows us to access the degree of spin purity in this branch also. This is illustrated in Fig. 4 (d) for the case of driving the spin-up sublevel. Phonon-assisted excitation to the upper branch is strongly suppressed at 4

K due to the large energy difference between the two branches for this centre. In the resulting spectra, shown in Fig. 4 (d), the transitions resonantly driven are indicated by red and blue arrows, respectively. Under these conditions, the contrast between fluorescence intensity originating from the two spin orientations (red and blue filled curve, respectively) is recovered to above 90%. This evidences high spin purity within both excited state branches. Therefore, by carefully selecting the driven transitions, spin-selective optical access to SiV$^-$ centres can be achieved in nanodiamonds as well as bulk diamond.

A natural extension of this work is to align the symmetry axis of a SiV$^-$ centre to the magnetic field either by rotating the samples or by implanting SiV$^-$ centres in a [111]-oriented diamond crystal. Both approaches will align the quantization axes of the ground and excited states and elicit the inherent optical selection rules linked to the spin orientation. All-optical ultrafast spin manipulation techniques or optically detected magnetic resonance can then give access to the electronic spin coherence of the pure ground state. Full coherent control of the SiV$^-$ spin state along with fluorescence-detection-based spin initialization and readout will be within reach for the realization of a highly efficient spin-photon quantum interface.

**Methods**

For the bulk diamond sample, SiV$^-$ centres were incorporated from residual silicon in the growth chamber during homoepitaxial CVD growth on a Ib high pressure high temperature diamond substrate (Sumitomo, (001) surface orientation), resulting in a thin (80-100 nm), high-quality layer with a homogeneous density of centres. Details concerning the growth conditions are described elsewhere [30]. Single SiV$^-$ centres in nanodiamonds were CVD grown on a silicon substrate, which was covered by an intermediate yttria-stabilized-zirconia buffer layer and atop a 150 nm iridium layer. Prior to growth, the substrate was seeded with de-agglomerated synthetic nanodiamonds with sizes up to 30 nm (Microdiamant Liquid Diamond MSY), where the seed density was chosen to result in a density of crystals containing SiV$^-$ of about one per 50 x 50 $\mu m^2$. Detailed growth conditions are given in [14]. The sample is mounted on a piezo-driven three-axis translation stage (attocube) in a helium bath cryostat (T=4.2 K) at the centre of a tunable 7-Tesla superconducting magnet (Cryogenics) in Faraday configuration. Excitation of the SiV$^-$ centres is performed non-resonantly at 700 nm (Coherent Mira-CW) and resonantly using a frequency-tunable external-cavity diode laser around 737 nm (Toptica DL Pro).  SiV$^-$ spectra were recorded using a spectrometer with ~40 $\mu eV$ resolution (PI Acton). All measurements are performed using a fibre-based confocal microscope with polarisation controlled excitation and collection (polarisers: Thorlabs LPVIS, half wave plates: Thorlabs AHWP05M-980). Light is focused onto the sample with a NA=0.68 aspheric lens (Thorlabs C330TME-B), and residual laser light in the detection arm is removed with either a 720 nm longpass filter (third millennium) or with a polariser set perpendicularly to the incoming laser polarisation. With the latter technique, laser suppression up to $5x10^{-7}$ was reached.

**Acknowledgements**

We gratefully acknowledge financial support by the University of Cambridge, the European Research Council (FP7/2007-2013)/ERC Grant agreement no. 209636, and FP7 Marie Curie Initial Training Network S$^3$NANO. We thank J. Maze, V. Waselowski, A. Gali, J. Becker, and C. Matthiesen for technical assistance and helpful discussions.

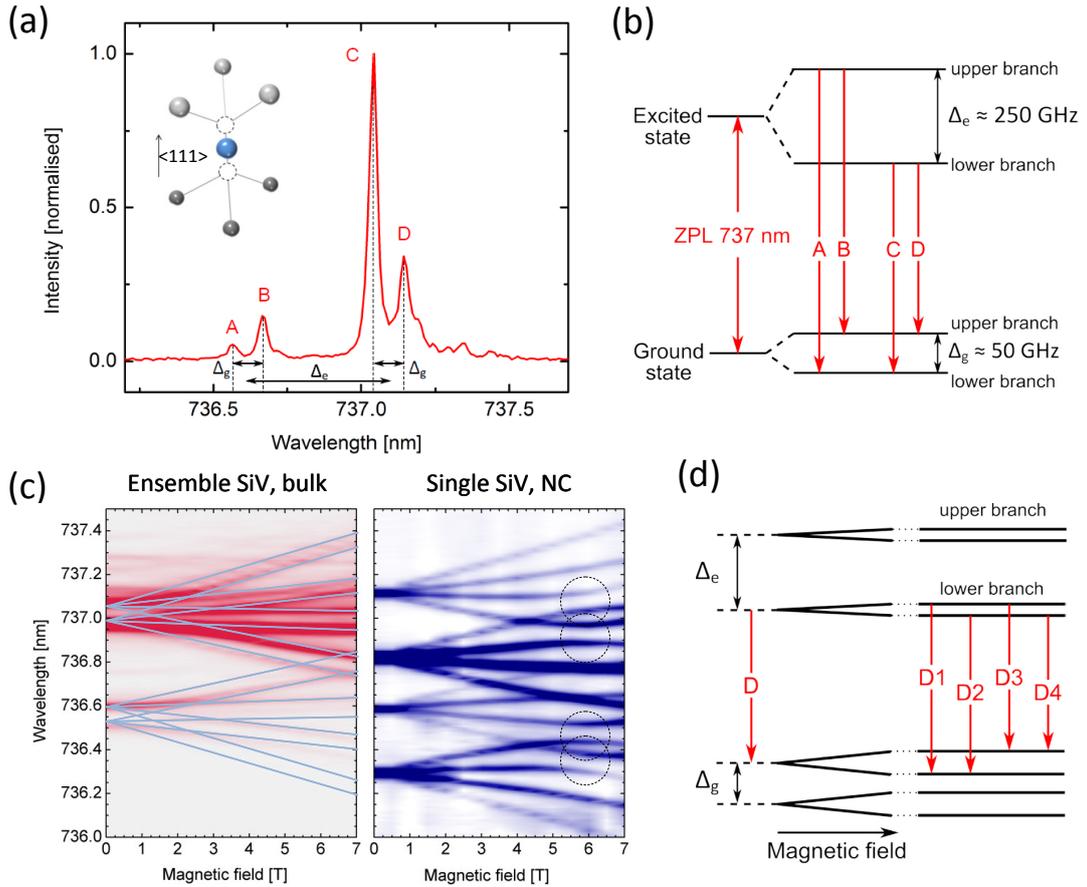

**Figure 1 – Introduction to the SiV⁻ centre**. (a) Fluorescence spectrum at 4 K for a SiV⁻ ensemble in [001]-oriented bulk diamond obtained by non-resonant excitation at 700 nm. The dominant transitions due to the ²⁸Si isotope, forming two doublets, are labelled from A to D, and the weaker, red-shifted transitions originate from ²⁹Si and ³⁰Si. The atomic structure of the SiV⁻ centre is shown in the inset, with the silicon atom (blue) in split-vacancy configuration between the unoccupied lattice sites (dashed circles) and the nearest neighbour carbon atoms (grey). (b) Energy level scheme of the SiV⁻ centre. The split ground and excited states give rise to four optical transitions, labelled from A to D according to the transitions in (a). (c) Fluorescence spectra at 4 K as a function of magnetic field along [001] for an ensemble of SiV⁻ centres in bulk diamond (red colourplot, light blue lines are a guide to the eye) and for a single SiV⁻ in a diamond nanocrystal (blue colourplot), where the strain induces a larger splitting of the doublets at zero field, resulting in a clearer splitting into a quadruplet for each transition. Avoided crossings in the spectrum, marked by black dashed circles, originate from spin-orbit coupling [27]. (d) Magnetic field splitting of each spin-1/2 energy level. The optical transitions split into four as shown for transition D due to the different g-factors in the ground and excited states.

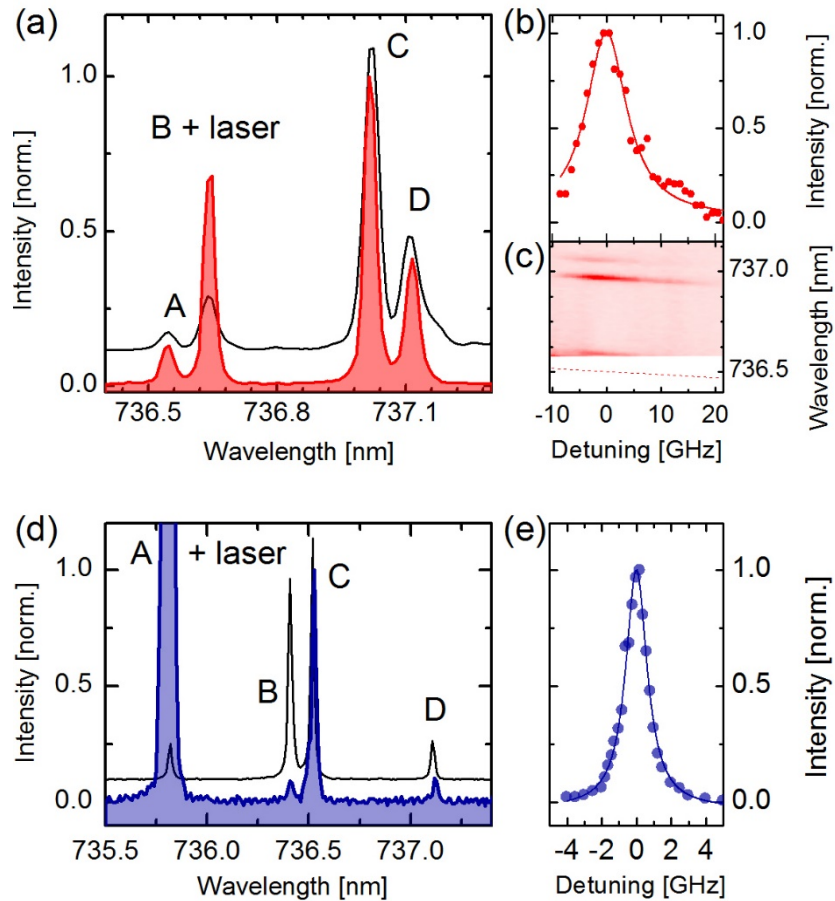

**Figure 2 – Resonance fluorescence at 0 T** (a) Resonance fluorescence spectrum at 4 K for a SiV⁻ ensemble (red shaded curve) with laser resonantly driving transition B, and non-resonant spectrum (black solid curve). (b) Photoluminescence excitation of transition A for a SiV⁻ ensemble. The maximum intensity of transition C is plotted as a function of the resonant laser detuning (red dots) and can be fitted by a single Lorentzian (red solid curve) with FWHM 9.1 ± 0.7 GHz. (c) As the laser is tuned towards lower wavelengths across transition A (red dashed line), the observed fluorescing transition wavelengths B - C exhibit the same blue shift in the spectrum. Consequently, for a given laser frequency only a sub-ensemble of the centres with identical strain conditions is excited. (d) Resonance fluorescence spectrum at 4 K for a single SiV⁻ in nanodiamond (confirmed via photon-correlation measurements), with laser resonantly exciting transition A (blue shaded curve), and non-resonant spectrum (black solid curve). The relative intensities of transitions under resonant excitation differ from the ones obtained in the non-resonant spectrum due to the polarisation-based laser suppression [22]. (e) Photoluminescence excitation of transition A for the same single centre in nanodiamond. The maximum intensity of peak C is plotted as a function of laser detuning (blue dots) and can be fitted by a single Lorentzian (blue solid curve) with FWHM of 1.4 ± 0.1 GHz.

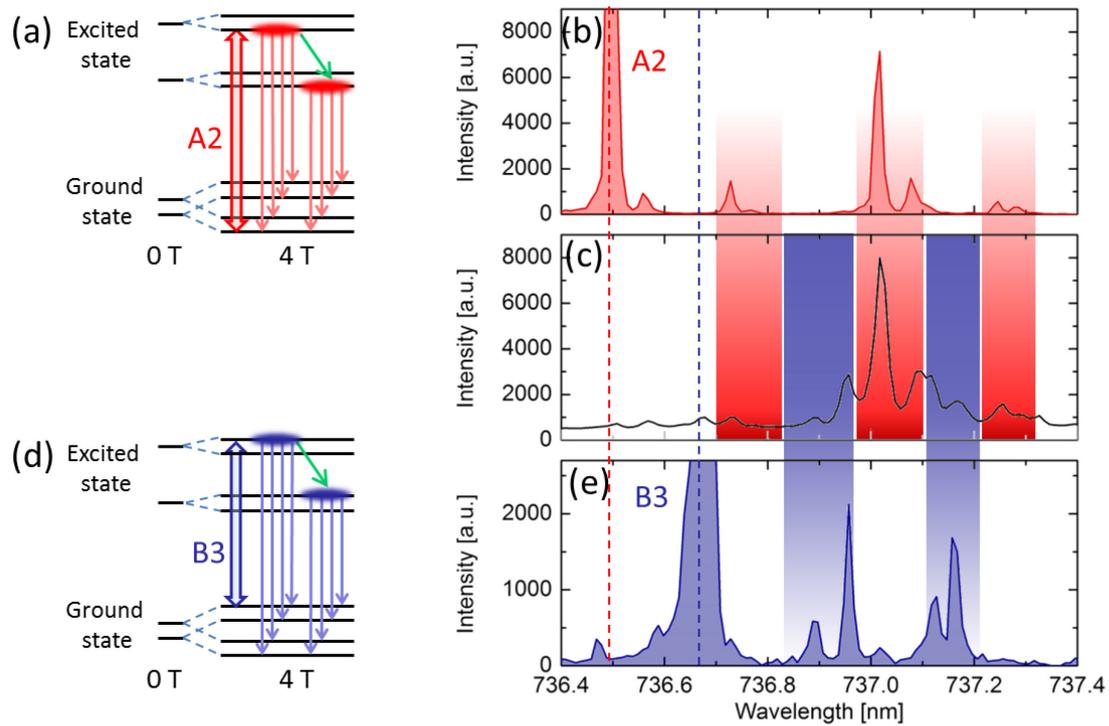

**Figure 3 – Resonance fluorescence of ensemble SiV⁻ at 4 T** (a) Energy level scheme for a SiV⁻ in bulk diamond at 0 T and 4 T. The driven transition populates the spin-down projection in the upper branch of the excited states and is indicated by a red double arrow. Relaxation takes place to the spin-down projection in the lower branch of the excited states as indicated by the green arrow. The thin red arrows indicate the transitions observed in the spectrum in panel (b). (b) Resonance fluorescence spectrum measured driving transition A2 at 4 T for a SiV⁻ ensemble in bulk diamond (red shaded curve). (c) Non-resonant fluorescence spectrum at 4 T for a SiV⁻ ensemble (black solid curve). (d) Energy level scheme for a SiV⁻ in bulk diamond at 0 T and 4 T. The driven transition populates the spin-up projection in the upper branch of the excited state and is indicated by a blue double arrow. Relaxation takes place to the spin-up projection in the lower branch of the excited state as indicated by the green arrow. The thin blue arrows indicate the transitions observed in the resonant spectrum [panel (e)]. (e) Resonance fluorescence spectrum measured when driving transition B3 at 4 T for a SiV⁻ ensemble (blue shaded curve). The fluorescing transitions in (b) and (e) are highlighted with red and blue panels respectively on the non-resonant spectrum in (c). The laser wavelengths used for resonant excitation in (b) and (e) are indicated with red and blue dashed lines, respectively.

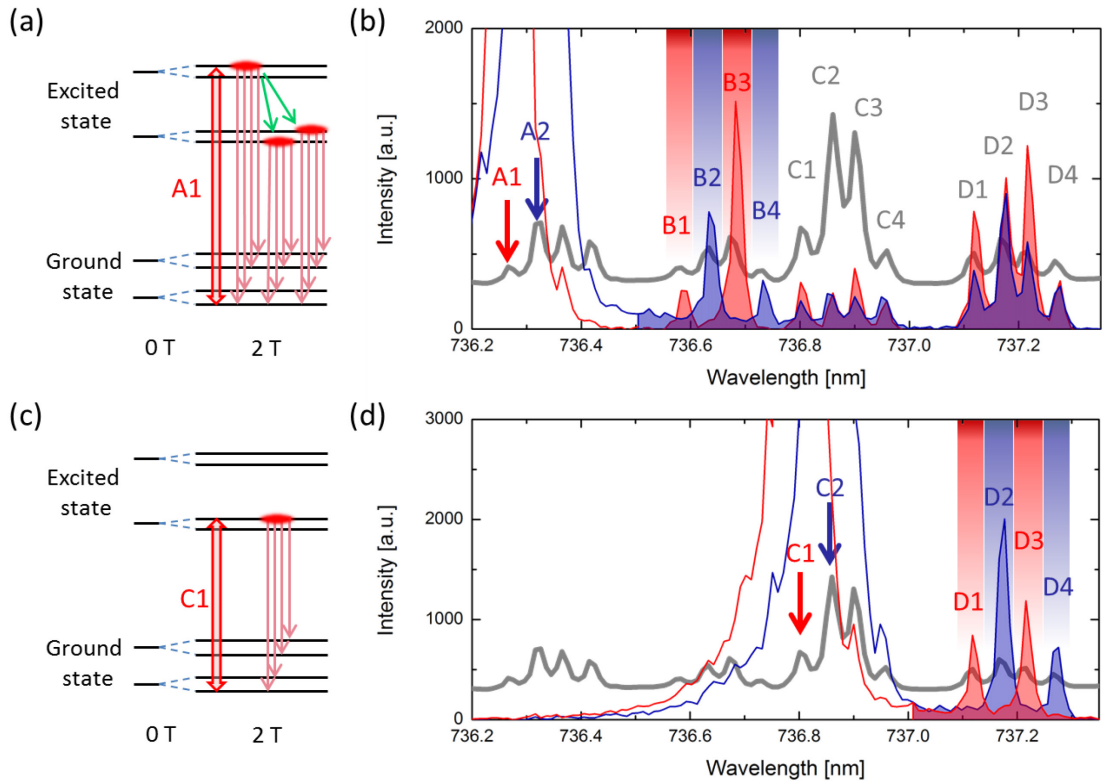

**Figure 4 – Resonance fluorescence at 2 T for a single SiV⁻ in nanodiamond**. (a) Energy level scheme for the single SiV⁻ in nanodiamond at 0 T and 2 T. Driving transition A1 (indicated by a red double arrow) results in populating the spin-up sublevel in the upper branch of the excited state. From there, relaxation takes place to both lower branch sublevels (green arrows). The spectrum obtained when driving this transition is shown in Panel (b) (red shaded curve), along with a spectrum when resonantly driving transition A2 (blue shaded curve). Driven transitions are indicated in the spectra by the red and blue arrows on the non-resonant spectrum (grey curve). (c) Same as panel (a), but when driving transition C1 (red double arrow), populating the spin-up sublevel in the lower branch. (d) The spectra obtained when driving C1 (red shaded curve) and C2 (blue shaded curve) along with the spectrum obtained under non-resonant excitation (grey curve).

14